\documentclass[sigconf]{acmart}
\usepackage{xcolor}
\usepackage{multirow}
\usepackage{stfloats}

\AtBeginDocument{%
  \providecommand\BibTeX{{%
    \normalfont B\kern-0.5em{\scshape i\kern-0.25em b}\kern-0.8em\TeX}}}

\copyrightyear{2025}
\acmYear{2025}
\setcopyright{cc}
\setcctype{by}
\acmConference[CHI '25]{CHI Conference on Human Factors in Computing Systems}{April 26-May 1, 2025}{Yokohama, Japan}
\acmBooktitle{CHI Conference on Human Factors in Computing Systems (CHI '25), April 26-May 1, 2025, Yokohama, Japan}\acmDOI{10.1145/3706598.3713189}
\acmISBN{979-8-4007-1394-1/25/04}

\begin{document}

\title{Development of the Critical Reflection and Agency in Computing Index}


\author{Aadarsh Padiyath}
\affiliation{%
  \institution{University of Michigan}
  \city{Ann Arbor}
  \country{USA}}
\email{aadarsh@umich.edu}

\author{Mark Guzdial}
\affiliation{%
  \institution{University of Michigan}
  \city{Ann Arbor}
  \country{USA}}
\email{mjguz@umich.edu}

\author{Barbara Ericson}
\affiliation{%
  \institution{University of Michigan}
  \city{Ann Arbor}
  \country{USA}}
\email{barbarer@umich.edu}


\begin{abstract}
As computing's societal impact grows, so does the need for computing students to recognize and address the ethical and sociotechnical implications of their work. While there are efforts to integrate ethics into computing curricula, we lack a standardized tool to measure those efforts, specifically, students' attitudes towards ethical reflection and their ability to effect change. This paper introduces the novel framework of Critically Conscious Computing and reports on the development and content validation of the Critical Reflection and Agency in Computing Index, a novel instrument designed to assess undergraduate computing students' attitudes towards practicing critically conscious computing. The resulting index is a theoretically grounded, expert-reviewed tool to support research and practice in computing ethics education. This enables researchers and educators to gain insights into students' perspectives, inform the design of targeted ethics interventions, and measure the effectiveness of computing ethics education initiatives.
\end{abstract}




\maketitle

\section{Introduction}

Computing's pervasive influence in society requires computing professionals to consider the ethical and sociotechnical implications of their artifacts \cite{ko2020time, vakil2018ethics, benjamin2019race, winner2017artifacts}.  However, scholars have critiqued our current approach to computing education, arguing it often neglects ethical and critical perspectives \cite{ko2020time, vakil2019s, vakil2018ethics, morales2023conceptualizing, raji2021you}. This neglect, they contend, has contributed to a culture of disengagement with ethics and the perception that critical perspectives are "not my job," irrelevant, or unimportant \cite{cech2014culture, padiyath2024realist, tran2024s, madaio2024learning, darling2024not}. In response, professional organizations such as ACM and IEEE have emphasized social impacts and public concerns in their codes of ethics \cite{gotterbarn1997software}, while accreditation bodies such as ABET have stressed the importance of legal and ethical considerations in post-secondary computing curricula \cite{ABET_2024}.

Interest in incorporating these issues into computing culture is growing, with many offering their own proposals \cite{morales2023conceptualizing}, theories \cite{padiyath2024realist}, and interventions \cite{brown2024teaching} on how to improve and evolve computing ethics pedagogy. While these approaches vary, a Freiran framework emphasizing the development of \textit{critical consciousness} is a compelling foundation \cite{freire2020pedagogy,ko2020time,ko2022critically, morales2023conceptualizing}. Scholars often conceptualize critical consciousness as \textit{critical reflection} -- involving a critical analysis of perceived inequalities and an endorsement of egalitarianism; \textit{critical agency} -- the perceived capacity and ability to effect social and political change; and \textit{critical action} -- individual and/or collective praxis \cite{watts2011critical, diemer2017development}.

However, post-secondary students' attitudes towards interventions aimed at instilling these ideals vary significantly, even within a single intervention \cite{padiyath2024realist, kirdani2022house, jarzemsky2023applies}. A recent review found students may be becoming more aware of the social impacts and consequences of computing, but often come to class with preconceived notions about these ideas \cite{padiyath2024realist, kirdani2022house, skirpan2018ethics, inclezan2014promoting}. Additionally, the review noted that although some students are readily accepting of ethics content, they were wary of applying ethical values in the workplace, believing they will lack agency or that there will be serious repercussions in such environments \cite{padiyath2024realist, castro2023piloting, dobesh2023towards, fiesler2021integrating, skirpan2018ethics}. As a result, they may eschew considering ethics and impacts as their responsibility \cite{padiyath2024realist, castro2023piloting, tran2024s}.

Although these efforts provide preliminary insights into the circumstances in which ethics pedagogy is effective and when it is not, there is limited empirical work on measuring students' ethical perspectives and beliefs \cite{brown2024teaching}. We need mechanisms for evaluating and keeping track of students' overarching views on ethical concepts. Rapid methods of evaluating understanding can guide the design and tailoring of interventions, identify opportunities to challenge students' preconceived notions, and promote growth.

To this end, this paper presents the development and content validation of the \textbf{Critical Reflection and Agency in Computing Index}: a set of two sub-scales designed to measure attitudes towards operationalizations of critical reflection and critical agency in computing. Guided by best practices for index development \cite{boateng2018best}, our main contributions are:

\begin{enumerate}
    \item Critically Conscious Computing: a conceptual framework for understanding operationalizations of critical reflection and agency in computing, developed through literature review and expert consultation, and

    \item The Critical Reflection and Agency in Computing Index: a rigorously developed heuristic survey with evidence for content and face validity that can be used to assess students' attitudes and beliefs in these constructs.
\end{enumerate}

This work enables concerted efforts to incorporate critical and ethical perspectives in computing education; identify patterns in students' ethical perspectives, inform curriculum and intervention design; enable longitudinal studies to monitor changes in attitudes over time; and compare across institutions to identify best practices in computing ethics education.
Developed for use as an initial heuristic tool, the Index draws on psychometric scale development principles while prioritizing immediate use. While future work will establish psychometric properties through refinement and quantitative validation, the current version of this tool helps educators and researchers to develop ethically informed and motivated computing professionals by providing an instrument to measure progress toward that goal.

\section{Background}

In the following section, we discuss the relevant background and prior work that informed the design and context of our index, while introducing our conceptualization of Critically Conscious Computing.

\subsection{Ethics in Computing Education}

For decades, the field of computing was broadly perceived as an apolitical, purely technical discipline \cite{vakil2018ethics, malazita2019infrastructures, benjamin2019race, ko2020time, ryoo2024show}. However, the last decade has seen a profound shift in this perspective, as educators, policy-makers, and the public increasingly recognize the social and ethical implications of technology \cite{o2017weapons, noble2018algorithms, eubanks2018automating, gray2019ghost, zuboff2019age, benjamin2019race, brown2024teaching, padiyath2024realist}. This evolution, catalyzed by high-profile tech scandals and the resulting "\textit{techlash}" \cite{su2021critical, sarder2022entering, kirdani2024taught, ryoo2024show}, has moved ethics from the sidelines to a core component of computing curricula \cite{ABET_2024, fiesler2020we}.

Traditionally, ethics education in computer science focused on individual decision-making, often through Western philosophical frameworks like utilitarianism, deontology, and social contract theory \cite{fiesler2020we}. Some programs adopted an embedded ethics model, which incorporates ethics content into technical courses \cite{grosz2019embedded}; however, this approach can sometimes result in prioritizing or "over-trusting" purely technical fixes to ethical problems, also called "techno-solutionism" \cite{morozov2013save, rogers2024lie}. This focus on Western or purely technical ethics-fixes has been critiqued for its limitations in addressing the complex, systemic nature of ethical challenges in computing \cite{raji2021you, vakil2018ethics, vakil2019s}.

In response, researchers in the field of Human-Computer Interaction (HCI) and Science \& Technology Studies (STS) have embraced more critical and transformative approaches to computing ethics. HCI and STS scholars often advocate for centering anti-racist, feminist, and explicitly political perspectives in computing ethics education \cite{abebe2022anti, bardzell2010feminist, keyes2019human, bardzell2016humanistic, costanza2020design}. This shift aligns with broader calls in the field to recognize computing as inherently value-laden and deeply embedded in structures of power and oppression \cite{winner2017artifacts, noble2018algorithms, benjamin2019race, vakil2019s, ko2020time, ko2022critically}. Such approaches demand sociotechnical understanding of how computing technologies interact with, influence, and embody power and politics \cite{winner2017artifacts, sengers2005reflective, bardzell2010feminist, vakil2019s, ko2020time, ko2022critically}.

The integration of ethics into computing education has thus progressed from simple awareness-raising to more complex discussions of social responsibility, criticality, and community engagement \cite{padiyath2024realist, ko2020time, yadav2022toward, brown2024teaching}. Early efforts in standalone computing ethics courses focused on helping students hone their ability to "issue-spot" and critique \cite{fiesler2020we}, often surprising them with the relevance of these concerns to their field \cite{padiyath2024realist}. However, recent observations suggest a potential shift in student attitudes, with many now entering computing classrooms interested and concerned about the ethical implications of technology, or with preconceived notions about such issues \cite{padiyath2024realist, padiyath2024undergraduate, fiesler2021integrating}.

Despite these encouraging signs, the exact nature and extent of this change in student attitudes remains unclear. We lack comparable data on students attitudes towards broader issues of ethics and criticality in computing, how attitudes differ across different populations, and whether these classroom discussions will transfer to ethical decision-making in the professional context \cite{brown2024teaching}. Without a specific, measurable understanding of students' attitudes and beliefs, we cannot know if we are preparing students properly for the ethical challenges they will face in the real world.

This shift challenges us to move beyond narrow technical considerations to engage with reflections of not just what can be built, but what should be built, and by whom \cite{bardzell2010feminist, vakil2019s, costanza2020design, garcia2020no, klassen2022run, benjamin2024imagination}. It calls for a reimagining of computing ethics education that addresses systemic issues and cultivates a sense of social responsibility and agency among future computing professionals \cite{vakil2018ethics, ko2020time, ko2022critically}.

\subsection{Critically Conscious Computing: A Novel Framework}

Building on these critical perspectives and drawing from Paulo Freire's theories of \textit{conscientização}, or critical consciousness \cite{freire2020pedagogy}, we propose the novel framework: \textbf{Critically Conscious Computing}. This framework synthesizes existing critical approaches and provides a structured way to conceptualize and assess ethical engagement in computing. We define Critically Conscious Computing as a form of computing practice in which individuals (1) analyze the sociotechnical implications of computing systems and practices, including issues of power, politics, culture, and equity (\textit{critical reflection}); (2) develop the sense of agency to question and challenge norms, assumptions, and practices in computing (\textit{critical agency}); and (3) take action to serve the needs and interests of diverse and marginalized communities, whether through creating new technologies, reshaping existing ones, or resisting the use of computing in harmful ways (\textit{critical action}).

From reflection to agency to action, critical consciousness theory posits that as individuals develop a deeper understanding of systemic issues and recognize their own capacity to effect change, they are more likely to engage in transformative praxis \cite{watts2011critical, diemer2017development}. This theoretical progression mirrors well-established frameworks in HCI, suggesting a solid theoretical foundation. For instance, the Security \& Privacy Acceptance Framework (SPAF) similarly emphasizes the importance of awareness (akin to critical reflection) alongside motivation and ability (related to agency) in influencing cybersecurity and privacy best-practice behaviors (actions) \cite{das2022security}. The Fogg Behavior Model \cite{fogg2019fogg}, frequently used for designing behavior change interventions, also aligns with our framework by highlighting the dynamic between prompts (which can stem from critical reflection) in relation to motivation and ability (agency) leading to successful behavior change (action). Further, the Theory of Planned Behavior \cite{ajzen1991theory}, used for explaining an individual's behaviors, posits that a person's attitudes and norms toward the behavior (reflection) alongside perceived control to perform a behavior (agency), influences intentions to perform the behavior. Critical consciousness theory extends these concepts to incorporate and hypothesize how ethical and critical perspectives transform into praxis.

\subsection{Designing for Critical Reflection and Agency in Computing Education}

Our definition of critically conscious computing is aspirational - a possible goal of critical CS pedagogy. However, many students may not have had the opportunity to deeply engage with the social, political, and ethical dimensions of computing, and may not be prepared to fully embody critically conscious computing \cite{padiyath2024realist, padiyath2024undergraduate, jarzemsky2023applies, kirdani2022house}. Students often enter the classroom with preconceived notions about technology's role in society, viewing it as a neutral tool divorced from social and ethical considerations \cite{padiyath2024realist, kirdani2022house}. Additionally, students often worry that "I will get fired if I speak up about ethics," believing they will have almost no agency in their workplace \cite{padiyath2024realist, castro2023piloting, widder2023s}. These preconceptions pose significant challenges and hurdles for educators and researchers designing interventions to foster critical consciousness~\cite{padiyath2024realist}.

Recognizing this, intervention designers are increasingly focusing on scaffolding students' development of critical reflection and agency \cite{rivera2024teaching, padiyath2024realist, widder2023s}. These approaches often prompt students to question dominant and uncritical narratives about computing \cite{kirdani2024identity} or recognize the human judgments inherent in their work \cite{brown2022shortest, brown2023designing}. Research shows that once students realize they are making value-laden decisions and can question prevailing narratives, they become more receptive to ethical reasoning and interventions \cite{padiyath2024realist}.

However, the effectiveness of these approaches can vary significantly across and within different contexts, interventions, and student populations \cite{padiyath2024realist}. Measuring this effectiveness remains a challenge for researchers and educators \cite{brown2024teaching}. Factors such as diverse student backgrounds and experiences, rapidly changing technological landscapes, and the breadth of topics covered in computing ethics all contribute to this complexity \cite{brown2024teaching}. We
lack 
a standardized tool to measure students' progress along this path, making it difficult to compare and contrast outcomes across different contexts or identify broader patterns and shifts in students' development of critical reflection and agency skills \cite{padiyath2024realist, brown2024teaching}.

To help address these challenges in designing effective interventions, we propose focusing on measuring alignment with foundational principles that serve as accessible gateways to critical reflection and agency in computing. These principles represent initial steps that students can engage with early in their computing education, laying the groundwork for developing a more comprehensive critically conscious computing practice over time. By assessing alignment with these foundational concepts, educators can better understand students' starting points and preconceived notions, tailor interventions more effectively, and track progress in developing a more critically conscious practice.

\subsection{Measuring Critical Reflection and Agency in Computing Education}

Current assessments of computing ethics education primarily rely on qualitative methods or narrowly focused surveys \cite{brown2024teaching, padiyath2024realist}. While these approaches offer valuable insights, they often are difficult to generalize and scale. Additionally, many existing assessments lack explicit grounding in theoretical frameworks, limiting their explanatory power \cite{padiyath2024realist, brown2024teaching}. 

Several validated instruments exist for measuring ethics and professional responsibility in engineering education, including the Test Of Ethical Sensitivity In Science And Engineering (TESSE) \cite{borenstein2008test}, the Engineering Professional Responsibility Assessment (EPRA) \cite{canney2016validity}, and the Student Engineering Ethical Development Survey (SEED) \cite{finelli2012assessment}. However, computing presents distinct ethical challenges. Software's rapid deployment, global reach, and its increasingly prevalent role in society create different ethical considerations than traditional engineering disciplines. Computing education researchers have noted these differences and called for computing-specific theories and tools \cite{al2016updated, nelson2018use}. Additionally, while discussions of practical professional responsibility (micro-ethics) are necessary, scholars in computing and engineering ethics emphasize the need to consider how technologies interact with our profession's systemic and collective responsibilities (macro-ethics) as well \cite{herkert2005ways, riley2008ethics, ko2020time, vakil2018ethics, benjamin2019race}. \citeauthor{horton2022embedding}'s Ethics Attitudes and Self-Efficacy (EASE) scale represents an important step in this direction, providing an approach to measuring computing students' general ethical attitudes before and after specific ethics interventions. Our work builds on this by integrating both micro-ethics (personal conduct and attitudes towards ethics in computing) and macro-ethics (systemic impacts of computing and systemic views of agency) into a unified expert-reviewed framework and heuristic tool for understanding students' broader development of critical consciousness in computing.

Recent efforts to measure critical consciousness more broadly, such as the Short Critical Consciousness Scale (ShoCCS), have made significant strides \cite{diemer2017development, diemer2022development}. However, these scales often target youth and thus lack the relevance and specificity required to assess critical consciousness in higher education and vocational contexts. Similarly, political efficacy scales, while relevant for measuring agency, mainly focus on making change in governmental contexts rather than professional environments \cite{craig1982measuring, caprara2009perceived, sarieva2018measure}.

Given that many interventions and proposals to incorporating ethics and critical pedagogy draw inspiration from Freire's concept of critical consciousness \cite{morales2023conceptualizing}, this theory provides a solid foundation for developing our scale. Additionally, we focus on computing due to its significant influence on contemporary culture and society \cite{benjamin2019race}, the broad perception that it is an apolitical field \cite{malazita2019infrastructures}, and the persistent lack of diversity among developers \cite{martin2018leaky}. As such, it would be beneficial to measure the attitudes of students towards these topics. In developing our scale, we also ground it in ideas recognized by professional agencies, such as those outlined in curricular guidelines \cite{10.1145/3664191} and in the Software Engineering Code of Ethics \cite{gotterbarn1997software}. This code explicitly discusses the necessity for the computing profession to engage with ethics and the social impacts of their work, aligning with our focus on critical reflection and agency.

A scale explicitly designed to measure and monitor critical consciousness in computing has the potential to unite and advance scholarship in this field. Therefore, this paper details the development of the Critical Reflection and Agency in Computing Index.

\section{Methods}

The development of our scale followed the process recommended by \citet{boateng2018best}. The authors detail three main steps to assure a rigorous scale -- (1) Domain Identification, (2) Item Generation, (3) and Gathering Evidence for Content Validity, which is mainly assessed through Expert Reviews and Cognitive Interviews. Our study protocol, including use of expert reviews and cognitive interviews, was approved by our university's Institutional Review Board (IRB).

To ground our work in existing scholarship, we conducted a review of recent computing ethics pedagogy literature. Closely following methods from a prior review in this area \cite{padiyath2024realist}, we searched the ACM Digital Library, Scopus, and Web of Science for work published in the last decade (2013-2023) using terms related to ethics, education, and computing. Selection criteria focused on: (1) articles discussing ethics integration in specifically computing education (excluding articles in engineering ethics), (2) papers presenting theoretical frameworks for critical/ethical computing, (3) studies measuring student attitudes toward ethics in computing, and (4) publications in peer-reviewed venues alongside magazine and position articles. We paid particular attention to literature discussing critical consciousness, ethical reflection, and student agency in computing contexts, analyzing how different approaches defined and operationalized these concepts. This review informed our domain identification and item generation process.

\subsection{Domain Identification}

We began with domain identification -- defining our constructs and their operationalization. This process was based on our literature review of computing ethics pedagogy research and consultation with content experts, as recommended by \citeauthor{mccoach2013instrument} \cite{mccoach2013instrument}. Our review led us to identify two foundational principles of critical reflection in computing: the non-neutrality of technology
-- appearing in several works such as \citeauthor{benjamin2019race} \cite{benjamin2019race}, \citeauthor{ko2020time} \cite{ko2020time}, and \citeauthor{vakil2018ethics} \cite{vakil2018ethics}, who all discuss how critical engagement with computing involves
recognizing that technologies embody the values of their creators and development contexts \cite{winner2017artifacts, vakil2018ethics, benjamin2019race, ko2020time, green2021data, yadav2022toward}; and the value of interdisciplinary insights
-- highlighted by \citeauthor{washington2020twice} \cite{washington2020twice} and \citeauthor{raji2021you} \cite{raji2021you} who note a rounded computing/ethics education involves 
emphasizing the importance of drawing on diverse expertise and epistemologies \cite{raji2021you, washington2020twice}. These align well with the broader definition of critical reflection: analyzing perceived inequalities and endorsing egalitarianism \cite{watts2011critical, diemer2017development}. While our review revealed less focus on agency in computing ethics education, we identified two key aspects of critical agency: the capacity to advocate for ethical practices within computing organizations
-- noted as important for acting on ethical problems by \citeauthor{widder2023s} \cite{widder2023s} and others
\cite{fiesler2020we, smith2023incorporating, rivera2024teaching}, and the belief in the responsiveness of computing institutions to ethical and critical concerns
-- emphasized in ethics interventions like \citeauthor{castro2023piloting}'s work \cite{castro2023piloting}, with \citeauthor{sarder2022entering} \cite{sarder2022entering} and \citeauthor{padiyath2024realist} \cite{padiyath2024realist} noting its role in student acceptance of ethics education
\cite{sarder2022entering, castro2023piloting, padiyath2024realist, madaio2024learning}. These aspects mirror conceptualizations of political efficacy \cite{craig1982measuring, miller1989american, groskurth2021english}: personal effectiveness and system responsiveness.

We operationalized critical reflection in two ways. First we focused on 
recurring concepts from articles often cited in our review:
including "Computing has limits" and "Data has limits" (from \citeauthor{ko2020time}'s critical CS education principles \cite{ko2020time}), and the notion that ethics discussions should center discussions of power (drawing from \citeauthor{vakil2019s}'s work \cite{vakil2019s}). Second, we adopted a broader perspective -- as recommended by \citeauthor{boateng2018best} to account for operationalizations outside of our theoretical frameworks -- as recognizing the need for explicit ethics and social impact discussions in computing training \cite{fiesler2021integrating}. To ground these in professional expectations, we drew items from the ACM/IEEE Software Engineering Code of Ethics \cite{gotterbarn1997software} and CS2023 Curricular Guidelines \cite{10.1145/3664191}. For critical agency, we focused on personal effectiveness as the belief in one's ability to uphold ethical conduct and communicate ethics perspectives (often the main goal of CS ethics courses \cite{fiesler2020we}), and system responsiveness as the belief that ethical concerns raised will be heard and addressed in computing projects and workplaces (theorized as necessary for enacting a critically conscious praxis \cite{watts2011critical, diemer2017development, diemer2022development, sarder2022entering, widder2023s}).

\begin{table*}
\centering
\begin{tabular}{p{0.15\textwidth}p{0.45\textwidth}p{0.23\textwidth}}
\toprule
\textbf{Domain} & \textbf{Construct} & \textbf{Operationalizations} \\
\midrule
\multirow{5}{*}{Critical Reflection} & 
\multirow{3}{*}{Recognizing computing/data embeds values/power} & 
Computing has limits \\

& & Data has limits \\

& & Centering power in ethics \\

& Recognizing computing training should include more explicit ethics and social impact discussions &  \\
\midrule
\multirow{2}{*}{Critical Agency} & 
\multirow{2}{*}{Belief that computing professionals have agency} & 
Personal effectiveness \\
& & System responsiveness \\
\bottomrule
\end{tabular}
\caption{Domains, Constructs, and Operationalizations of Critical Reflection and Agency in Computing}
\label{table:critical_reflection_agency}
\end{table*}

\subsection{Item Generation}

Following the domain identification, the first author developed an initial pool of 45 items to measure the constructs. 
For transparency, these items are presented in Appendix \ref{appendix:initialitems}. 
Our development process was guided by \citeauthor{fowler1995improving}'s essential characteristics for quality survey items which emphasize the items should be kept simple, straightforward, and should follow the conventions of normal conversation \cite{fowler1995improving, boateng2018best}.

The first author created items for each of the operationalized constructs, in addition to items related to computing professionals' expected responsibilities (another of \citeauthor{ko2020time}'s critical CS education principles \citeauthor{ko2020time}) and the need for non-siloed discussions of computing's impacts (based on work by \citeauthor{raji2021you} \cite{raji2021you}). However, these were later subsumed into a different construct or removed during the expert review process to improve clarity and avoid redundancy. To ensure comprehensive and accurate coverage of the constructs, we drew from three sources: (1) Original items created based on our operationalizations, (2) adaptations of examples from previous work detailing the operationalizations, and (3) modified versions of existing scale items -- a similar process used to create items for the original critical consciousness scales \cite{diemer2017development, diemer2022development}. For critical reflection, half the items are reverse-coded to measure criticality through a rejection of uncritical views and mitigate response bias. The item pool was iteratively refined through discussions among the authors, focusing on clarity and brevity. We chose a 4 or 6-option Likert scale response format (Disagree/Agree), balancing nuanced responses with rapid assessment needs, aligning with previous critical consciousness measures like ShoCCS \cite{diemer2022development}.

\subsection{Expert Review}

After item generation, we conducted an iterative external expert review process to assess and refine the content validity of our scale \cite{boateng2018best, iglesias2016reporting}. We engaged seven experts with diverse backgrounds in computing ethics, critical pedagogy, and survey design. These experts were identified as persons with "in-depth knowledge of the topic of interest gained through their life experience, education, or training" \cite{iglesias2016reporting}. 
Given our goal to develop the scale for educational contexts, we focused our expert review on those with direct experience in computing ethics and critical pedagogy.
These included two PhD Candidates studying student attitudes towards computing ethics and critical pedagogy in CS classrooms, four professors specializing in ethics and justice-centered computing pedagogy, and a consultant with expertise in questionnaire item design and development. All our experts (excluding the questionnaire expert) have recently published peer-reviewed academic research regarding computing ethics or critical pedagogy in computing or engineering education research venues.

We provided each expert with the construct definitions, operationalizations, and item pool. The review process was structured using an Excel spreadsheet format, where items were listed alongside their respective constructs and operationalizations. Experts were asked to qualitatively evaluate each item based on the following criteria: (1) sufficient coverage of the construct, (2) appropriateness for measuring said construct, (3) accuracy and completeness, (4) clarity and difficulty level, and (5) any other issues or concerns. Reviewers provided feedback both synchronously and asynchronously. During \textasciitilde 30 minute synchronous sessions, the first author took notes, while for asynchronous feedback, reviewers recorded their evaluations directly in the Excel spreadsheet.

This process was conducted iteratively, beginning with the graduate student reviewers, followed by the survey development expert, and concluding with professor content experts for gathering evidence for content validity. After each round of feedback, we modified or removed problematic items accordingly. This process aligns with both best practices outlined by \citeauthor{boateng2018best}, as well as practices employed by HCI studies regarding survey development \cite{Bentvelzen2021development, borgert2023home}.

\subsection{Cognitive Interviews}

To refine our scale and ensure its appropriateness for our target population, we conducted cognitive interviews with 5 undergraduate computer science majors -- within the number of participants recommended by \citeauthor{boateng2018best} \cite{boateng2018best}. Students were recruited from 
beginner and intermediate-level 
programming courses at a large public R1 university in the United States
to verify the scale was accessible to students at earlier stages of their education.
Students were offered an incentive of \$20 to participate.
To avoid possible social desirability bias and conflict of interest, students did not have any previous relationship with the researcher conducting the interviews.
We employed a think-aloud protocol \cite{boateng2018best}, asking participants to verbalize their thought processes while responding to each item. Follow-up questions were asked to gain deeper insights into their interpretation of the questions when necessary. These focused on items that participants found confusing or too difficult to answer. Online interviews were conducted, recorded, and transcribed using Zoom videoconferencing software, while the first author took notes during in-person interviews. After each interview, the scale was modified if necessary, including rewording items for clarity. This process continued until we reached a stable version of the scale with minimal issues -- as recommended by best practices \cite{boateng2018best}.

\subsection{Limitations}

This study, while rigorous in its approach to index development, has several limitations. First, our expert review panel, while diverse, was limited in size and 
consisted primarily of academic professionals. Although this aligned with our initial focus on educational applications, future versions of the scale would benefit from industry practitioner perspectives.
Additionally, our cognitive interviews were conducted with students from a single institution, potentially limiting the generalizability of our findings. Future work could involve testing and refining the index with a larger sample
and expanding the expert review panel to include both academic and industry voices.

The Critical Reflection and Agency in Computing Index, like all measurement tools, has inherent limitations \cite{diemer2023illustrating, frisby2024critical}. While we aimed to develop a comprehensive measure, the complex and nuanced nature of critical consciousness in computing means that some aspects may not be fully captured. The use of Likert-scale items, while efficient, does not allow for the in-depth understanding afforded by qualitative methods. Additionally, self-report measures are subject to social desirability bias -- the tendency of survey respondents to answer in a manner viewed favorably by others \cite{chung2003exploring} -- especially when dealing with issues of ethics \cite{tan2021social}. 
Finally, while our development of this index followed best practices for scale development and content validation, it is important to note that we have yet to conduct a full validation study. We present the Index as a heuristic tool -- rather than a fully validated psychometric instrument -- to enable immediate practical application while being transparent about its current stage of development. This means the Index can effectively guide discussions and inform intervention design, but users should interpret quantitative results as preliminary indicators. Future validation work will establish statistical properties such as factor structure, reliability, and refinement, allowing for more quantitative applications of the Index.

In line with the critical approach that informs our work, we must also acknowledge the inherent limitations of quantitative measures in capturing complex social phenomena, a perspective known as CritQuant \cite{diemer2023illustrating, frisby2024critical}. Quantitative scales, including ours, risk oversimplifying multifaceted concepts and reinforcing dominant power structures by determining what is measured and how. Further, we recognize the act of measurement itself is not neutral and can shape the phenomena being studied. While we believe our index provides valuable insights, we encourage users to complement it with qualitative methods and remain aware of its limitations and potential unintended consequences. We view this index not as a definitive measure, but as a tool to facilitate discussions and further research in realizing a more critical computing education.

\section{Results}

\subsection{Expert Reviews}

Expert reviews yielded unanimous agreement on the conceptual definitions and operationalizations of the constructs, validating the index's theoretical foundation. However, several issues emerged during the review process. The survey design expert noted the scale's ambitious number of subconstructs, prompting significant restructuring. This included consolidating items and removing of two subscales: "recognizing computing professionals have responsibilities", and "recognizing discussion of computing's impact should not be siloed". These concepts were subsumed as individual items within the other critical reflection subscales.

Experts suggested some improvements for clarity and recommended replacing jargon with more common synonyms to improve student understanding. The survey design expert emphasized using the phrase "Please indicate the extent to which you \textit{agree or disagree} with the following items," as opposed to our original wording of "... you agree with the following items," to mitigate potential bias in question wording.

One expert highlighted criticality as a "moving target," suggesting the need to consider how the questions and our broader cultural context might evolve over time. This prompted discussions about potential future iterations of the index to maintain relevance, though no concrete solution was reached.

Another expert suggested providing more context for the critical agency questions to account for various power dynamics, such as "I will promote an ethical practice in my workplace, even if initially met with resistance from my \textit{peers}" and "I will promote an ethical practice in my workplace, even if initially met with resistance from my \textit{supervisor}." However, we decided against this to maintain the scale's broad usability and focus on capturing their attitudes towards their own personal effectiveness, rather than asking about where their values apply and don't apply. Alternative scales measuring this construct could incorporate vignettes to capture a more nuanced understanding of critical agency.

As the review process progressed, a strong consensus emerged noting the scale's overall quality and potential impact. One expert appreciated our approach grounding the index in theory, noting it provides explanatory power and suggests clear pedagogical actions. Another expert expressed interest in administering the index in her course the following week. This positive reception, coupled with the constructive feedback received, affirmed the robustness of our development process and the usefulness of the index.

\subsection{Cognitive Interviews}

Cognitive interviews with five undergraduate computing students identified certain phrases and wordings that students initially struggled with. For example, a student noted she was unsure of the meaning of, "Requirements building for software." 

\begin{quote}
    "\textit{Requirements} building for software [\textit{pause}], requirements \textit{building} for software [\textit{pause}], hmm, this one is confusing."
\end{quote}

When the item was explained to her, she noted "Identifying requirements to build software" as clearer phrasing. This modification was implemented, with no further issues with this item reported in other interviews.

One participant, despite feeling her peers' often disregard ethics considerations, initially indicated that she didn't know more about computing ethics compared to other software developers. This led to modifying the item to specify "I am better informed about the ethics and societal impacts of technology than most of my software developer \textit{peers}" to more accurately capture the intended personal effectiveness.

Interestingly, participants often considered various contexts when responding to critical agency items, yet still provided generalized answers. For example:

\begin{quote}
    "I have never had to talk about ethical computing issues with my peers. I think my friends would listen, but my classmates might not. So I'll do agree."
\end{quote}

\begin{quote}
    "I think about working at a company as a younger dev, I think I would just say, 'ok!' [in response to a request to develop a problematic product]. But I wouldn't always sit back if there was something blatantly wrong." <selects disagree>
\end{quote}

This aligns with our goal of capturing broader attitudes rather than responses to specific scenarios, while also suggesting the potential benefit of a more nuanced critical agency survey in the future.

The interviews provided evidence for face validity of our constructs. For example, students who expressed less confidence in their ability to communicate about ethics consistently marked their personal effectiveness lower. Similarly, those who believed not everyone needs to consider ethical issues tended to rate non-technical computing training considerations lower. For instance, one student noted:

\begin{quote}
    "I don't know if everyone should have to know about social impacts." <selects disagree> ... "I think we should leave legal considerations to a lawyer." <selects disagree>
\end{quote}

Some interviewees acknowledge uncertainty about workplace responsiveness to ethical issues, as they lacked actual software development professional experience. However, these students still marked their responses based on their beliefs, which aligns with the survey's intent to capture individual beliefs about these operationalizations.

\begin{quote}
    "I can't really speak to computing specifically, since I haven't had an internship yet, but based on workplaces in the past, there's places to go that are devoted to ethical issues." <selects agree>
\end{quote}

After the second interviewee, the next three participants reported no difficulty understanding or answering the questions, apart from considering their own responses. Therefore, we believe saturation was reached in this case, providing sufficient evidence for the survey's face validity.

\subsection{Final Index}

After incorporating feedback from expert reviews and cognitive interviews, our final Critical Reflection and Agency in Computing Index consists of 40 items distributed across two main constructs: Critical Reflection (30 items) in Tables \ref{tab:critical_reflection1} and \ref{tab:critical_reflection2} and Critical Agency (10 items) in Table \ref{tab:critical_agency}. Each construct is further divided into sub-constructs to capture different aspects of critical consciousness in computing. Each sub-construct within the "Recognizing computing/data embeds values/power" operationalization includes both positively- and reverse-coded items to mitigate response bias. We recommend administering the index with a 4- or 6-item Likert-scale format, with respondents indicating their level of agreement or disagreement with each statement.

\begin{table*}
    \centering
    \begin{tabular}{p{2.5cm}p{3.5cm}p{7.5cm}}
    \toprule
    \multicolumn{3}{p{13.5cm}}{\textbf{Question Wording}: Computing technologies have wide-ranging impacts on society. Please indicate the extent to which you agree with the following statements:} \\
    \midrule
        \textbf{Construct} & \textbf{Operationalization} & \textbf{Item} \\
    \midrule
        \multirow{11}{8em}{Recognizing computing/data embeds values/power:} & \textit{Computing has limits} & Computing should inform, not replace, human decision-making. \\
        & \textit{Computing has limits} & It’s important to explore if a non-computing solution would be the best approach for a given problem. \\
        & \textit{Computing has limits} & With enough resources, computing technologies can solve any problem. (R) \\
        & \textit{Computing has limits} & We should prioritize computational solutions over human judgment. (R) \\
        & \textit{Data has limits} & Data reflects the past and does not fully capture current reality. \\
        & \textit{Data has limits} & Using data requires considering the data's limitations. \\
        & \textit{Data has limits} & Datasets that are large enough can overcome any bias in collection. (R) \\
        & \textit{Data has limits} & Biases in datasets can always be corrected with the right techniques. (R) \\
        & \textit{Centering power in ethics} & Considering issues of social justice should be a fundamental consideration in the design and development of any computing system. \\
        & \textit{Centering power in ethics} & Developing computer software for public use requires input from marginalized groups. \\
        & \textit{Centering power in ethics} & Ethics discussions in computing should only involve computer scientists. (R) \\
        & \textit{Centering power in ethics} & Computing technologies benefit everyone equally. (R) \\
    \bottomrule
    \end{tabular}
    \caption{Critical Reflection Subscale: Recognizing computing/data embeds values/power. (R) denotes a reverse-coded item.}
    \label{tab:critical_reflection1}
\end{table*}

\begin{table*}
    \centering
    \begin{tabular}{p{3cm}p{10.5cm}}
    \toprule
    \multicolumn{2}{p{13.5cm}}{\textbf{Question Wording}: Different professions require different training. Please indicate the extent to which you agree that the following should be part of training for every software engineer:} \\
    \midrule
        \textbf{Construct}   & \textbf{Item} \\
    \midrule
    \multirow{18}{8em}{Recognizing computing training should include more explicit ethics and social impact discussions:} & The social impacts of software. \\
     & The environmental impacts of software. \\
     & Legal considerations in software development. \\
     & Engaging with stakeholders affected by software projects. \\
     & Ethical implications of topics being studied. \\
     & Collaborating on software development projects with local community groups. \\
     & Software development professional responsibilities. \\
     & Guidelines for discussing ethical issues with others. \\
     & A software development code of ethics. \\
     & Discrete mathematics. (0) \\
     & Computer architectures. (0) \\
     & Databases. (0) \\
     & Technical programming skills. (0) \\
     & Delivering software projects on-time and within budget. (0) \\
     & Software quality assurance and testing. (0) \\
     & Computer science theory and algorithms. (0) \\
     & Identifying requirements to build software. (0) \\
     & Data structures. (0) \\
    \bottomrule
    \end{tabular}
    \caption{Critical Reflection Subscale: Recognizing computing training should include more explicit ethics and social impact discussions. (0) denotes a comparison item.}
    \label{tab:critical_reflection2}
\end{table*}

\begin{table*}
    \centering
    \begin{tabular}{p{3cm}p{3cm}p{7.5cm}}
    \toprule
    \multicolumn{3}{p{13.5cm}}{\textbf{Question Wording}: Please indicate the extent to which you agree with the following:		} \\
    \midrule
        \textbf{Construct} & \textbf{Operationalization} & \textbf{Item} \\
    \midrule
    \multirow{10}{10em}{Belief that computing professionals have agency:} & \textit{Personal effectiveness} & I have a good understanding of the important ethical and social impacts to consider when developing software. \\
    & \textit{Personal effectiveness} & I am able to participate in discussions about ethics and social impacts of computing. \\
    & \textit{Personal effectiveness} & I am confident in my own ability to uphold ethical conduct in software development. \\
    & \textit{Personal effectiveness} & I am better informed about the ethics and societal impacts of technology than most of my software developer peers. \\
    & \textit{Personal effectiveness} & When working on computing projects with others, I could effectively voice my perspectives on ethical issues. \\
    & \textit{Personal effectiveness} & I will promote an ethical practice in my workplace, even if initially met with resistance. \\
    & \textit{System responsiveness} & There are processes within workplaces to handle reported ethical computing violations or concerns. \\
    & \textit{System responsiveness} & When I talk about ethical computing issues, my peers usually pay attention. \\
    & \textit{System responsiveness} & Software development professionals are allowed to have a say about ethical computing concerns at their workplaces. \\
    & \textit{System responsiveness} & When ethical computing concerns are raised by employees, workplaces are responsive to addressing these concerns. \\
    \bottomrule
    \end{tabular}
    \caption{Critical Agency Subscale}
    \label{tab:critical_agency}
\end{table*}

\section{Discussion}

The Critical Reflection and Agency in Computing Index, grounded in the Critically Conscious Computing framework, offers a novel tool for assessing ethical awareness and agency among computing students. This section discusses the implications of our work and considers its connection to broader issues in critical pedagogy research.

\subsection{Designing for Critically Conscious Computing}

This index provides educators and researchers with a heuristic tool to inform the design and evaluation of ethics interventions and courses in computing education. By using the index as a pre- and post-survey tool, educators can tailor their approaches based on students' initial attitudes and measure the effectiveness of their interventions. For instance, students who score low on the "computing has limits" items may benefit from targeted discussions about the limitations of computing and the importance of human-centered design approaches. Similarly, low scores on the "personal effectiveness" subscale could indicate the need for more practice in ethical decision-making scenarios and interventions to boost students' confidence in communicating their values.

The standardized nature of our index supports the development of best-practices and theory-building in CS ethics pedagogy. For example, if students consistently show higher critical agency scores after a particular intervention, that intervention could be highlighted and potentially adapted for use in other institutions. This approach to pedagogy development could advance our knowledge in a field that has often relied on narrow surveys or case studies.

Beyond individual classrooms, the index enables consideration of students' long-term trajectories in developing critical consciousness. By administering the survey throughout a students' education, we can track how attitudes evolve from introductory computing courses to graduation and even into early career stages. This longitudinal perspective could help us note where interventions might be most impactful or necessary.

While developed for educational contexts, the Index has potential applications in other research settings. HCI researchers studying design justice could use the framework and tool to evaluate how different design processes, tools, or practices influence students' critical consciousness in computing. While our content validation focused on educational contexts with students and we have not yet tested the Index in professional settings, the framework and tool may also be valuable in training contexts where practitioners are developing their critical consciousness in computing. However, using the Index to assess organizational culture or practitioners' existing beliefs would require additional validation work with industry experts and professional participants.

\subsection{Considerations for Critical Computing Pedagogy Research}

The Critically Conscious Computing framework emphasizes the relationship between technical skills and the development of critical reflection and agency. This challenges researchers and educators to design learning experiences that simultaneously develop technical proficiency, awareness of the limitations and ethical considerations around their work, and the sense of agency to act on their ethical values.

The focus on critical agency specifically addresses an under-theorized aspect of computing ethics education. While much work has focused on awareness-raising and fostering critical reflection, less attention has been paid to developing students' sense of agency in addressing ethical issues \cite{padiyath2024realist}. Our index provides both a conceptual framework and a measurement tool to explore this critical component, offering new research on how to effectively cultivate agency alongside sociotechnical skills.

The development of this index also highlights several important areas for future research. For example, exploring how students' backgrounds, experiences, and cultural contexts influence their development of critical consciousness in computing; investigating the trajectory of critical reflection development beyond recognizing the human decisions in their work; and examining the relationship between our index and actual ethical decision-making and behavior in professional settings.

\section{Conclusion}

This paper presents the development of the Critical Reflection and Agency in Computing Index, a novel measurement tool grounded in our Critically Conscious Computing framework. Through a methodologically rigorous approach, we have created an instrument to better understand computing students' perspectives on the ethical and social implications of technology. Our index distills complex critical consciousness concepts into expert-reviewed measurable constructs with evidence for face and content validity. This article offers two contributions: (1) a theoretical framework for Critically Conscious Computing that synthesizes existing critical approaches in computing education and HCI, and (2) a practical tool to systematically assess and measure the development of critical reflection and agency among computing students. These contributions provide researchers and educators with both a conceptual foundation and a concrete means to measure progress in fostering ethical awareness and agency. As the field continues to grapple with the ethical challenges posed by technologies, this index provides a valuable means to measure progress and inform the design of educational interventions. This work represents a step towards focusing our efforts to better prepare students for ethical and socially responsible computing practice.

\begin{acks}
These work was created with funding from a SIGCSE Special Projects Grant.
\end{acks}

\bibliographystyle{ACM-Reference-Format}
\bibliography{sample-base}

\clearpage
\appendix

\section{Initial Item Pool}
\label{appendix:initialitems}

The following tables present the initial pool of items generated during the development of our index. These represent our initial operationalizations of the constructs prior to expert reviews and cognitive interviews. We provide these items for transparency of our scale development process.

\begin{table*}[b]
    \centering
    \begin{tabular}{p{10cm}p{4.5cm}p{1cm}}
    \toprule
    \multicolumn{3}{p{13.5cm}}{\textbf{Question Wording}: Computing technologies have wide-ranging impacts on society. Please indicate the extent to which you agree with the following statements:} \\
    \midrule
        \textbf{Item} & Item adapted from & Citation \\
    \midrule
        Computing should augment, not replace, human decision-making. & "myth: software is always right" & \cite{ko2020time} \\
        It’s important to examine critically whether a non-computing solution would be the best approach for a given problem. & "software can only solve some problems, and many cases, creates new ones." & \cite{ko2020time} \\
        With enough development, computing technologies have the potential to solve virtually any problem. (R) & "myth: software can solve every problem" & \cite{ko2020time} \\
        We should always prioritize computational solutions over human judgment. (R) & "myth: software is always right" & \cite{ko2020time} \\
        & & \\
        Data reflects the past and may not fully capture current reality. & "data is always about the past and not the future" & \cite{ko2020time} \\
        Responsible data use requires considering the data's limitations and biases. & Written by first author &  \\
        Datasets that are large enough can overcome any bias in collection. (R) & "data is always an imperfect and biased record" & \cite{ko2020time} \\
        Biases in datasets can be corrected with the right "debiasing" techniques. (R) & "data is always an imperfect and biased record" & \cite{ko2020time} \\
        & & \\
        Addressing issues of social justice should be a fundamental consideration in the design and development of computing systems. & "We must highlight how technologies used to facilitate and automate our daily activities can lead to further racialization and injustices." & \cite{yadav2022toward} \\
        Ethical computing requires engaging with perspectives of marginalized groups. & "attending to how computing systems intersect with structures of inequality and hierarchy in society" & \cite{vakil2019s} \\
        Computing technologies benefit everyone equally. (R) & Written by first author &  \\
    \bottomrule
    \end{tabular}
    \caption{Initial Item Pool: Recognizing computing/data embeds values/power.}
    \label{tab:initialitems1}
\end{table*}

\begin{table*}
    \centering
    \begin{tabular}{p{5cm}p{7cm}p{3.5cm}}
    \toprule
    \multicolumn{3}{p{13.5cm}}{\textbf{Question Wording}: Different professions require different training. Please indicate the extent to which you agree that the following should be part of training for every software engineer:} \\
    \midrule
         \textbf{Item} & Item adapted from & Citation \\
    \midrule
      The social impacts of software engineering. & "Approve software only if they have a well-founded belief that it is safe, meets specifications, passes appropriate tests, and does not diminish quality of life, diminish privacy or harm the environment. The ultimate effect of the work should be to the public good." & SWE CoE \#1.03 \cite{gotterbarn1997software} \\
      The environmental aspects of software engineering. & "Identify, define and address ... environmental issues related to work projects." & SWE CoE \#3.03 \cite{gotterbarn1997software} \\
      Legal considerations in software engineering practice. & "Identify, define and address ... legal ... issues related to work projects." & SWE CoE \#3.03 \cite{gotterbarn1997software}  \\
      Engaging with stakeholders affected by software projects. & "Cooperate in efforts to address matters of grave public concern caused by software, its installation, maintenance, support or documentation." & SWE CoE \#1.05 \cite{gotterbarn1997software} \\
      Requirements building for software systems. (0) & "Knowing how to build something is of little help if we do not know what to build." & CS2023: Software Engineering, "SE-Requirements" \cite{10.5555/3664191.C5954055} \\
      Technical programming skills. (0) & Written by first author &  \\
      How to deliver software projects on-time and within budget. (0) & Written by first author &  \\
      Software quality assurance and testing. (0) & "Understand the role of testing, failure modes, and differences between good tests and poor ones." & CS2023: Software Engineering, "SE-Validation" \cite{10.5555/3664191.C5954055} \\
      Computer science theory and algorithms. (0) & "Explain the role of algorithms for writing programs." & CS2023: Software Development Fundamentals, "SDF-Algorithms" \cite{10.5555/3664191.C5954054} \\
      Artificial intelligence. (0) & "Determine when an AI approach is appropriate for a given problem, identify appropriate representations and reasoning mechanisms, implement them, and evaluate them..." & CS2023: Artificial Intelligence, "AI-Introduction" \cite{10.5555/3664191.C5954043} \\
      Data structures. (0) & "... Be able to select and use appropriate data structures..." & CS2023: Software Development Fundamentals, "SDF-DataStructures" \cite{10.5555/3664191.C5954054} \\
    \bottomrule
    \end{tabular}
    \caption{Initial Item Pool: Recognizing computing training should include more explicit ethics and social impact discussions.}
    \label{tab:initialitems2}
\end{table*}

\begin{table*}
    \centering
    \begin{tabular}{p{6.5cm}p{6.5cm}p{2.5cm}}
    \toprule
    \multicolumn{3}{p{13.5cm}}{\textbf{Question Wording}: Different professions have different responsibilities. Please indicate the extent to which you agree that the following are part of the professional responsibilities of software engineers:} \\
    \midrule
        \textbf{Item} & Item adapted from & Citation\\
    \midrule
         Reflecting on their own priorities, values, and perspectives which may influence the technologies they create. & "Temper all technical judgments by the need to support and maintain human values." & SWE CoE \#4.01 \cite{gotterbarn1997software}\\
         Protecting the health and safety of the public. & "In all these judgments concern for the health, safety and welfare of the public is primary."  & SWE CoE Preamble \cite{gotterbarn1997software}  \\
         Continuously learning about the social implications of their work. & "Software engineers shall participate in lifelong learning regarding the practice of their profession and shall promote an ethical approach to the practice of the profession." & SWE CoE \#8 \cite{gotterbarn1997software} \\
         Mitigating ethical issues in software development projects. & "Identify, define and address ethical ... issues related to work projects." & SWE CoE \#3.03 \cite{gotterbarn1997software}\\
         Addressing social justice issues in software development products. & Written by first author &   \\
         Delivering functional software that meets specified requirements. (0) & Ensure that specifications for software on which they work have been well documented, satisfy the users’ requirements and have the appropriate approvals. & SWE CoE \#3.08 \cite{gotterbarn1997software} \\
         Identifying, defining and addressing legal issues related to work projects. & "Identify, define and address ... legal ... issues related to work projects." & SWE CoE \#3.03 \cite{gotterbarn1997software} \\
         Balancing employer’s/organizational goals with ethical and social considerations. & "Moderate the interests of the software engineer, the employer, the client and the users with the public good." & SWE CoE \#1.02 \cite{gotterbarn1997software} \\
    \bottomrule
    \end{tabular}
    \caption{Initial Item Pool: Recognizing computing professionals have responsibilities.}
    \label{tab:initialitems3}
\end{table*}

\begin{table*}
    \centering
    \begin{tabular}{p{7.5cm}p{7cm}p{1cm}}
    \toprule
    \multicolumn{3}{p{13.5cm}}{\textbf{Question Wording}: Software engineering often involves solving complex problems that often have large social impacts. Please indicate the extent to which you agree that software engineers should do the following when working on projects:} \\
    \midrule
         \textbf{Item} & Item adapted from & Citation \\
    \midrule
        Seek expertise of non-software engineers to better understand the problem space. & "we recommend to focus on thinking and acting differently by including broad non-CS expertise and researchers when dealing with technical artifacts which have clear social impact." & \cite{raji2021you} \\
        Seek input from community representatives to understand the needs of diverse stakeholders. & "Demonstrate examples of effective collaborative outcomes to students in the form of ... advocacy campaigns done in conjunction with affected communities..." & \cite{raji2021you} \\
        Seek advice of external software engineers to gain different perspectives on problem-solving. (0) & Written by first author &  \\
        Seek guidance from software engineer colleagues within their organization to efficiently solve problems. (0) & Written by first author &  \\
        Seek direction from their organization/employer to ensure alignment with business objectives. (0) & Written by first author &  \\
    \bottomrule
    \end{tabular}
    \caption{Initial Item Pool: Recognizing discussion of computing’s impacts should not be siloed.}
    \label{tab:initialitems4}
\end{table*}

\begin{table*}
    \centering
    \begin{tabular}{p{7.5cm}p{5cm}p{3cm}}
    \toprule
    \multicolumn{3}{p{13.5cm}}{\textbf{Question Wording}: Please indicate the extent to which you agree with the following:		} \\
    \midrule
     \textbf{Item} & Item adapted from & Citation \\
    \midrule
    I have a pretty good understanding of the important ethical and social issues to consider when engineering software. & "I feel that I have a pretty good understanding of the important political issues facing our country." & Internal Political Efficacy Scale \cite{niemi1991measuring} \\
    I am well qualified to participate in discussions about ethics and social impacts of computing. & "I consider myself to be well qualified to participate in politics." & Internal Political Efficacy Scale \cite{niemi1991measuring} \\
    I am confident in my skills to uphold ethical conduct in software engineering. & Written by first author &  \\
    I am better informed about the ethics and societal impacts of technology than most software engineers. & "I think that I am better informed about politics and government than most people." & Internal Political Efficacy Scale \cite{niemi1991measuring} \\
    When working on computing projects, I can effectively voice my perspectives on ethical practices. & "I have the confidence to take active part in a discussion about political issues." & Political Efficacy Short Scale \cite{groskurth2021english} \\
    I will promote an ethical practice in my workplace, even if initially met with resistance. & Written by first author &  \\
    & & \\
    I believe there are processes within computing institutions to handle reported ethical violations or concerns. & "There are many legal ways for citizens to successfully influence what the government does" & External Efficacy Item \cite{craig1982measuring, scotto2021alternative} \\
    I believe reports of ethical concerns in computing would be properly addressed by relevant institutions. & Written by first author &  \\
    I feel computing institutions would be responsive to upholding ethical standards if concerns are raised. & Written by first author &  \\
    \bottomrule
    \end{tabular}
    \caption{Initial Item Pool: Belief that computing professionals have agency.}
    \label{tab:initialitems5}
\end{table*}



\end{document}